\documentclass[aps,prb,fleqn,12pt,superscriptaddress]{revtex4}
\usepackage{epsfig,color}
\usepackage{graphicx}
\usepackage{pdfpages}
\usepackage{braket}
\usepackage{amssymb,amsmath}
\usepackage{hyperref}
\begin{document}
\title{Spin-orbital order and excitations in $3d^4$, $4d^4$, and $5d^4$ systems: \\
Application to {\boldmath $\rm BaFeO_3$}, {\boldmath $\rm Sr_2RuO_4$}, {\boldmath $\rm Sr_2YIrO_6$}, and {\boldmath $\rm K_2OsCl_64$} }
\author{Shahid Ahmad}
\affiliation{Department of Physics, Indian Institute of Technology, Kanpur - 208016, India}
\author{Harshvardhan Parmar}
\affiliation{Department of Physics, Indian Institute of Technology, Kanpur - 208016, India}
\author{Shubhajyoti Mohapatra}
\affiliation{Government Autonomous College, Rourkela - 769004, Odisha, India}
\author{Avinash Singh}
\email{avinas@iitk.ac.in}
\affiliation{Department of Physics, Indian Institute of Technology, Kanpur - 208016, India}
\date{\today} 
\begin{abstract}
Evolution of composite spin-orbital order and coupled spin-orbital excitations is studied in a variety of $d^4$ systems with $n$=$4$ electrons in the $t_{2g}$ orbital sector using the generalised self-consistent + fluctuations approach for a realistic interacting-electron model. Within this unified approach, applications are discussed to compounds with $3d$, $4d$, $5d$ transition-metal ions such as $\rm BaFeO_3$ ($\rm Fe^{4+}$), $\rm Ca_2RuO_4$ ($\rm Ru^{4+}$), $\rm Sr_2RuO_4$ ($\rm Ru^{4+}$), $\rm Sr_2YIrO_6$ ($\rm Ir^{5+}$), and $\rm K_2OsCl_6$ ($\rm Os^{4+}$). Continuous interpolation from the nominally $S=1$ antiferromagnetic order in $\rm Ca_2RuO_4$ (strong crystal field, intermediate spin-orbit coupling (SOC) and Coulomb interaction) to the $J=0$ state relevant for $5d^4$ compounds (strong SOC, weak Coulomb interaction) and to the half-metallic ferromagnetic order when crystal field is negligible as in $\rm BaFeO_3$ (weak SOC, strong Coulomb interaction) and $\rm Sr_2RuO_4$ (intermediate SOC and Coulomb interaction) provides new fundamental insights into the magnetism of spin-orbit coupled systems. 
\end{abstract}
\maketitle
\newpage

\section{Introduction}

Due to the rich interplay between spin-orbit coupling (SOC), crystal field, and Coulomb interaction, the perovskite-structured $d^4$ compounds having 4 electrons in the $t_{2g}$ sector ($d_{yz}$, $d_{xz}$, $d_{xy}$ orbitals) of the octahedrally coordinated $\rm MO_6$ transition-metal ions exhibit a remarkably diverse set of ground states. 

While the layered perovskite compound $\rm Ca_2RuO_4$ (with highly distorted $\rm RuO_6$ octahedra and intermediate values of Hubbard interaction $U \sim 2$ eV, Hund’s coupling $J_{\rm H}\sim U/5$, and SOC value $\lambda \sim 150$ meV for the $4d$ element Ru) exhibits an insulating state at low temperature with nominally $S=1$ canted antiferromagnetic (AFM) order,\cite{nakatsuji_JPSP_1997,kunkemoller_PRL_2015,porter_PRB_2018} the same structured compound $\rm Sr_2RuO_4$ shows quasi two-dimensional Fermi liquid behaviour below 30 K temperature.\cite{mackenzie_RMP_2003,mravlje_PRR_2011,stricker_PRL_2014} 



On the other hand, the cubic perovskite compound $\rm BaFeO_3$ (with stronger $U\sim 3$ eV and weaker SOC $\lambda \sim 30$ meV for the $3d$ element Fe) exhibits a half-metallic ground state with strong ferromagnetic (FM) order.\cite{hoedl_JPCC_2022} Half-metallic FM state was also reported in $\rm Sr_2RuO_4$ from density functional theory (DFT) investigations where the FM order was suggested to be unstable due to spin fluctuations.\cite{huang_SCIREP_2020} Whereas, the double perovskite compounds such as $\rm Sr_2YIrO_6$, $\rm Ba_2YIrO_6$, and the cubic antifluorite type halides $\rm K_2OsCl_6$, $\rm K_2OsBr_6$ involving the $5d^4$ ions of $\rm Ir^{5+}$ and $\rm Os^{4+}$ (with weaker $U \sim 1$ eV and stronger SOC $\lambda \sim 500$ meV) exhibit the nominally total angular momentum $J=0$ ground state with zero spin and orbital moments.\cite{khaliullin_PRL_2013,cao_PRL_2014,corredor_PRB_2017,dey_PRB_2016,fuchs_PRL_2018,kusch_PRB_2018,nag_PRB_2018,hoshin_JMMM_2018,zhang_PRB_2022,warzanowski_PRB_2023}

It is interesting to note that the $5d^4$ compounds $\rm Sr_2OsO_4$, $\rm Ca_2OsO_4$, $\rm Ba_2OsO_4$ in the layered perovskite category do not exist, and the cubic perovskite $5d^4$ compounds $\rm CaOsO_3$, $\rm SrOsO_3$, $\rm BaOsO_3$ are formed at extremely high pressure,\cite{shi_JACS_2013,mohitkar_DALTRANS_2018} suggesting an intriguing role of presence of magnetic moments and magnetic interactions in stabilising the crystal structure. 

A unified understanding of how these different ground states observed in the $d^4$ compounds emerge and evolve from each other in different parameter regimes will be helpful in highlighting the rich interplay between SOC, crystal field, and Coulomb interaction. Towards this end, we will investigate the composite spin-orbital order and excitations within the generalised self-consistent + fluctuations approach for a realistic three-orbital model for the $d^4$ compounds. This unified approach has been recently employed for studying compounds with different electron numbers in the $t_{2g}$ sector, such as $\rm Sr_2 IrO_4$ ($n=5$), $\rm Ca_2 RuO_4$ ($n=4$), $\rm Na OsO_3$ ($n=3$), $\rm Sr_2 CrO_4$ ($n=2$), and $\rm Sr_2 VO_4$ ($n=1$).\cite{mohapatra_JPCM_2020,mohapatra_JPCM_2021,mohapatra_JPCM_2023_I,mohapatra_JPCM_2023_II} 

Two important features of the rich interplay found\cite{mohapatra_JPCM_2020} in $\rm Ca_2RuO_4$ --- (i) strong Coulomb enhancement of SOC and (ii) magnetic reorientation transition from AFM (planar) order to FM ($z$) order with decreasing crystal field --- will be shown to be profoundly significant in determining electronic and magnetic properties of the $d^4$ compounds, such as the electronic band structure and Fermi surface in the half-metallic state of $\rm Sr_2RuO_4$ and the excitation energy of the spin-orbit exciton across the renormalised spin-orbit gap in the $J=0$ state of the $5d^4$ compounds. Interestingly, the predicted magnetic transition in $\rm Ca_2RuO_4$ has been very recently confirmed in photo-excitation based studies.\cite{li_NATPHYS_2025}

In the case of the $n=4$ compound $\rm Ca_2 RuO_4$, a rich variety of collective excitations were obtained including magnon, orbiton, spin-orbiton, spin-orbit exciton, and high-energy out-of-phase magnon.\cite{mohapatra_JPCM_2021} It will be fascinating to see how these excitations evolve from $\rm Ca_2 RuO_4$ to $\rm Sr_2RuO_4$, and compare the calculated excitation energies with recent resonant inelastic X-ray scattering (RIXS) measurements in $\rm Sr_2RuO_4$.\cite{fatuzzo_PRB_2015,suzuki_NATCOM_2023} In the $J=0$ state, the collective excitations are expected to evolve to well-defined, propagating spin-orbit exciton modes, and a similar comparison with recent RIXS measurements\cite{kusch_PRB_2018,warzanowski_PRB_2023} of excitation energies in the $5d^4$ compounds should be of significant interest. 

The structure of this paper is as below. The model Hamiltonian and methodology is briefly reviewed in section II. After discussing earlier works on the $5d^4$ compounds in section III and on $\rm Sr_2RuO_4$ in section V, calculated results for the composite spin-orbital order and coupled spin-orbital excitations are presented in these two sections for the $J=0$ state and the half-metallic FM state, respectively. The instability of the $J=0$ state for SOC strength below the critical value, as inferred from the extrapolated exciton energy turning negative, is discussed in section IV. Finally, the unified picture of how the order and excitations evolve with SOC, crystal field, and Coulomb interaction, as realised in the different $3d^4$, $4d^4$, $5d^4$ systems, is discussed and concluded in section VI. 

\section{Model Hamiltonian and Methodology}
We will consider the Hamiltonian ${\cal H} = {\cal H}_{\rm band} + {\cal H}_{\rm cf} + {\cal H}_{\rm int} + {\cal H}_{\rm SOC}$ within the $t_{\rm 2g}$ manifold. The three-orbital ($\mu=d_{yz}, d_{xz}, d_{xy}$), two-spin ($\sigma=\uparrow,\downarrow$) basis is defined with respect to the common spin-orbital coordinate axes oriented along the planar M-O-M directions where M refers to the transition metal ion. The Hamiltonian terms have been discussed earlier,\cite{mohapatra_JPCM_2020} and are briefly reviewed below.

The hopping terms $t_1$, $t_2$, $t_3$ for the $d_{xy}$ orbital correspond to first, second, and third neighbours, respectively, while $t_4$ and $t_5$ are the first-neighbour hopping terms for the $d_{yz}$ ($d_{xz}$) orbital in $y$ $(x)$ and $x$ $(y)$ directions, respectively. The crystal field splitting is represented by the energy offset $\epsilon_{xy}$ for the $d_{xy}$ orbital relative to the degenerate $d_{yz}/d_{xz}$ orbitals. We have taken hopping parameter values: ($t_1$, $t_2$, $t_3$, $t_4$, $t_5$)=$(-1.0, 0.5, 0, -1.0, 0.2)$ unless otherwise mentioned, all in units of the realistic hopping energy scale $|t_1|$, which will be mentioned below for individual compounds.   


For the bare spin-orbit coupling term, we consider (for lattice site $i$): 
\begin{eqnarray} 
{\cal H}_{\rm SOC} (i) & = & -\lambda {\bf L} \cdot {\bf S} = -\lambda (L_z S_z + L_x S_x + L_y S_y) \nonumber \\ 
&=& \sum_\alpha \begin{pmatrix} \psi_{\mu \uparrow}^\dagger & \psi_{\mu \downarrow}^\dagger \end{pmatrix} \begin{pmatrix} i \sigma_\alpha \lambda /2 \end{pmatrix} 
\begin{pmatrix} \psi_{\nu \uparrow} \\ \psi_{\nu \downarrow} 
\end{pmatrix} + {\rm H.c.}
\label{soc}
\end{eqnarray}
in the spin-space representation, where the spin component $\alpha=z$ for the orbital pair $(\mu,\nu)=(d_{yz}, d_{xz})$, and similarly for other components. Since spin rotation symmetry is explicitly broken, the SOC term therefore generates anisotropic magnetic interactions from its interplay with other Hamiltonian terms. In the following, we will consider bare SOC values ($\lambda$) in the range 25-50 meV, 100-150 meV, and 400-500 meV for the $3d$, $4d$, and $5d$ compounds, respectively. 

Finally, for the Coulomb interaction terms, we consider (for site $i$):
\begin{eqnarray}
{\cal H}_{\rm int} (i) &=& U\sum_{\mu}{n_{i\mu\uparrow}n_{i\mu\downarrow}} + U^{\prime \prime}\sum_{\mu<\nu} n_{i\mu} n_{i\nu} - 2J_{\mathrm H} \sum_{\mu<\nu} {\bf S}_{i\mu}.{\bf S}_{i\nu} +J_{\mathrm P} \sum_{\mu \ne \nu} a_{i \mu \uparrow}^{\dagger} a_{i \mu\downarrow}^{\dagger}a_{i \nu \downarrow} a_{i \nu \uparrow} 
\label{h_int}
\end{eqnarray} 
in the $t_{2g}$ basis ($\mu,\nu=d_{yz}, d_{xz}, d_{xy}$). Here $U$ and $U^{\prime \prime}=U-5J_{\rm H}/2$ denote the intra-orbital and inter-orbital density interaction terms, while $J_{\rm H}$ and $J_{\rm P}=J_{\rm H}$ are the Hund's coupling and pair hopping interaction terms. Also, $a_{i\mu \sigma}$ and $a_{i\mu\sigma}^{\dagger}$ are the electron annihilation and creation operators for site $i$, orbital $\mu$, spin $\sigma=\uparrow ,\downarrow$. In terms of the electron field operator $\psi_{i\mu}^\dagger=(a_{i\mu\uparrow}^{\dagger} \; a_{i\mu\downarrow}^{\dagger})$, the spin density operator ${\bf S}_{i\mu} = \psi_{i\mu}^\dagger ${\boldmath $\sigma$}$ \psi_{i\mu}$, density operator $n_{i\mu\sigma}=a_{i\mu\sigma}^\dagger a_{i\mu\sigma}$, and total density operator $n_{i\mu}=n_{i\mu\uparrow}+n_{i\mu\downarrow}=\psi_{i\mu}^\dagger \psi_{i\mu}$. All interaction terms above possess continuous spin rotation symmetry as they are invariant under the corresponding SU(2) transformation. The $U$ values will be taken as $\approx$ 3eV, 2eV, and 1eV for the $3d,4d,5d$ compounds, respectively. 

As mentioned earlier, we will use the generalized self consistent + fluctuations approach wherein all orbital diagonal and orbital mixing condensates of the spin and charge operators ($\langle \psi_{i\mu}^{\dagger} \makebox{\boldmath $\sigma$} \psi_{i\nu} \rangle$ and $\langle \psi_{i\mu}^{\dagger} {\bf 1} \psi_{i\nu} \rangle$) are self-consistently determined in the Hartree-Fock (HF) approximation of the Coulomb interaction terms. The generalized self consistent approach in the two-sublattice basis allows for staggered and entangled orbital orders as well as ferromagnetic and antiferromagnetic spin orders.

The orbital mixing charge and spin condensates yield the orbital moments and Coulomb interaction induced SOC renormalization: 
\begin{eqnarray}
\langle L_\alpha \rangle & = & -i \left [ \langle \psi_\mu^\dagger \psi_\nu\rangle - \langle \psi_\mu^\dagger \psi_\nu\rangle^* \right ] = 2\ {\rm Im}\langle \psi_\mu^\dagger \psi_\nu\rangle \nonumber \\
\lambda^{\rm int}_\alpha & = & (U'' - J_{\rm H}/2)\  {\rm Im}\langle \psi_\mu^\dagger \sigma_\alpha \psi_\nu\rangle
\label{phys_quan}
\end{eqnarray}
where the orbital pair ($\mu,\nu$) corresponds to the component $\alpha=x,y,z$ as given below Eq. \ref{soc}. The last equation yields the Coulomb renormalised SOC components $\lambda_\alpha = \lambda + \lambda_\alpha^{\rm int}$ where $\lambda$ is the bare SOC value. The imaginary parts of orbital mixing condensates correspond to SOC induced orbital and spin orbital currents.

To investigate spin-orbital fluctuations, we will consider the time-ordered generalized fluctuation propagator:
\begin{equation}
[\chi({\bf q},\omega)] = \int dt \sum_i e^{i\omega(t-t')} 
e^{-i{\bf q}.({\bf r}_i -{\bf r}_j)} 
\times \langle \Psi_0 | T [\sigma_{\mu\nu}^\alpha (i,t) \sigma_{\mu'\nu'}^{\alpha'} (j,t')] |\Psi_0 \rangle 
\end{equation}
in the self-consistent ground state $|\Psi_0 \rangle$. The orbital mixing terms $\psi_\mu ^\dagger \makebox{\boldmath $\sigma$}\psi_\nu$ and $\psi_\mu ^\dagger \psi_\nu$ for the spin and charge operators are included here since the corresponding condensates were included in the generalized self consistent approach. At lattice sites $i,j$, the generalized spin-charge operators: $\sigma_{\mu\nu}^\alpha = \psi_\mu ^\dagger \sigma^\alpha \psi_\nu$ include both orbital diagonal ($\mu=\nu$) and mixing ($\mu\ne\nu$) cases. With $\sigma^\alpha$ defined as Pauli matrices for $\alpha=x,y,z$ and unit matrix for $\alpha=c$, both spin and charge operators are included. 

Evaluation of the generalized fluctuation propagator in the random phase approximation (RPA) has been recently discussed along with application to several compounds with electron fillings $n=1$ to 5 in the $t_{\rm 2g}$ sector.\cite{mohapatra_JPCM_2021,mohapatra_JPCM_2023_I,mohapatra_JPCM_2023_II} The spectral function of the fluctuation propagator:
\begin{equation}
{\rm A}_{\bf q}(\omega) = \frac{1}{\pi} {\rm Im \; Tr} [\chi({\bf q}, \omega)]_{\rm RPA}
\label{spectral}
\end{equation}
provides information about the collective spin-orbital excitations (magnon, orbiton, spin-orbiton), where the character is determined from the basis resolved contributions in the composite $\mu\nu\alpha$ basis. Orbiton and spin-orbiton modes correspond to same-spin and spin-flip particle-hole excitations, respectively, involving different orbitals. 


\begin{figure}[t]
\hspace*{0mm}
\psfig{figure=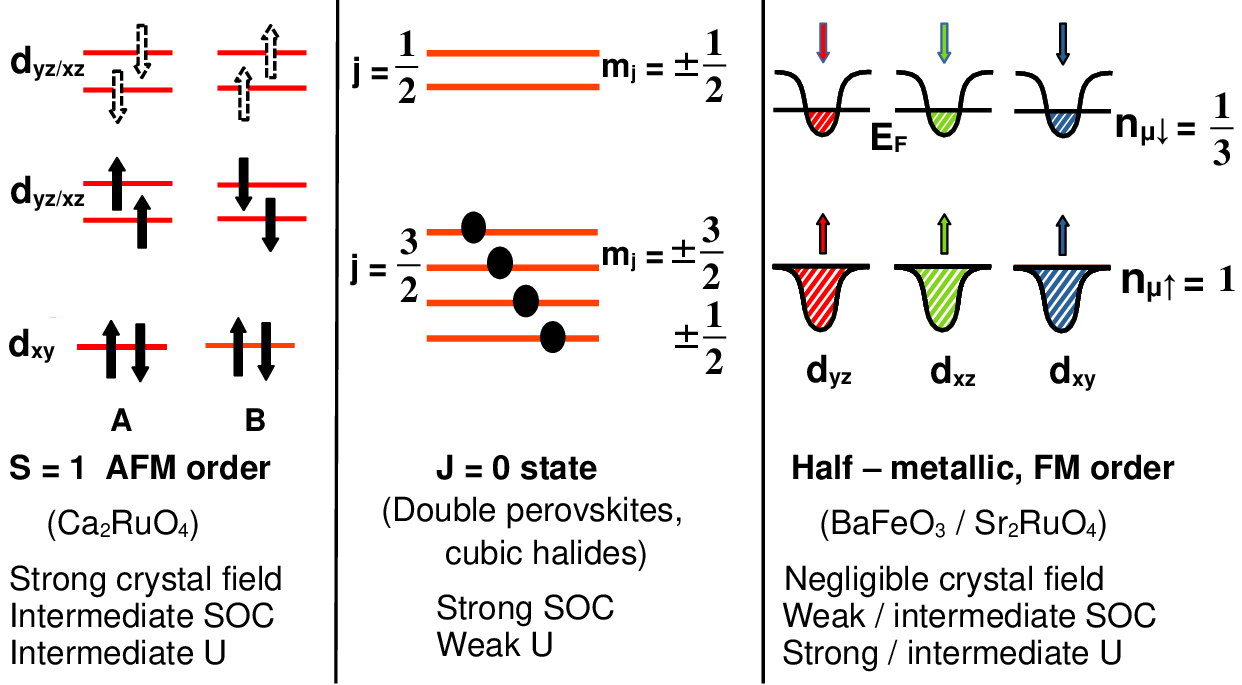,angle=0,width=120mm}
\vspace{0mm}
\caption{Schematic diagram showing different magnetic states of the $3d^4$, $4d^4$, $5d^4$ systems with $n=4$ electrons in the $t_{2g}$ sector depending on strength of SOC, crystal field, and Coulomb interaction.} 
\label{fig1}
\end{figure} 

In the following, we will consider as a reference case the (i) nominally $S=1$ AFM order in $\rm Ca_2RuO_4$ (strong crystal field $\epsilon_{xy} \sim -1$, intermediate SOC)  and investigate evolution to two other parameter regimes. Since Hund’s coupling is antagonistic to SOC induced spin-orbital entanglement, we will show that reducing $J_{\rm H}/\lambda$ leads to the (ii) $J=0$ state relevant for the $5d^4$ systems such as double perovskites $\rm Sr_2YIrO_6$ and cubic halides $\rm K_2OsCl_6$. On the other hand, for weak SOC and negligible crystal field ($\epsilon_{xy} \sim 0$), the (iii) half-metallic FM order is obtained with electron densities $n_\mu^\uparrow \sim 1$ and $n_\mu^\downarrow \sim 1/3$, so that $n_\mu \sim 4/3$ and the local moments $m_\mu \sim 2/3$ for all three nearly degenerate orbitals $\mu=d_{yz},d_{xz},d_{xy}$, as observed in the $3d^4$ cubic compound $\rm BaFeO_3$. In the itinerant electron description, the emergent FM spin couplings between local moments is generated by exchange of the particle-hole pair. 

The spin-orbital orders for the three cases discussed above are schematically shown in Fig. \ref{fig1}. In the $4d^4$ layered perovskite compound $\rm Sr_2RuO_4$, we will see that long range FM order is unstable when transverse spin fluctuations are considered. In the following, we present our results for the evolution of spin-orbital order and excitations between these different parameter regimes ($\lambda,\epsilon_{xy},U,J_{\rm H}$). We will consider a uniform layered perovskite lattice structure in order to focus on the continuous interpolation. Dominantly local features such as the high-energy spin-orbital excitations should provide good approximation for the $5d^4$ double perovskite and cubic halide systems in which the transition metal ions form a face centered cubic lattice. 
 
\section{$J=0$ state: Application to $\rm 5d^4$ systems}

Initial experimental studies of $5d^4$ double perovskites such as $\rm Sr_2YIrO_6$ reported weak paramagnetic behaviour and magnetic moment $\mu_{\rm eff}$=0.91$\mu_{\rm B}$, challenging theoretical expectation of the non-magnetic $J=0$ state.\cite{cao_PRL_2014} Magnetic order below 1.3 K was also reported, possibly linked to a non-cubic crystal field. However, later studies attributed the weak paramagnetic response to low concentration (0.4-0.7\%) of impurities due to weak chemical disorder, with no evidence of long-range magnetic order between 1.8 K and 300 K.\cite{corredor_PRB_2017} The measured magnetic susceptibility $\sim 5.90 \times 10^{-4}$ emu/mol and corresponding  magnetic moment $\mu_{\rm eff} = 0.21 \mu_{\rm B}$ were much smaller than expected for a nominally $S=1$ system. 

Similarly, experiments on $\rm Ba_2YIrO_6$ revealed a faint magnetic signal which was ascribed to correlated magnetic moments with effective moment $\mu_{\rm eff}=0.44\mu_{\rm B}$ per Ir atom, and no sign of long-range magnetic order was found down to 0.4 K.\cite{dey_PRB_2016,fuchs_PRL_2018} This cubic compound was initially considered to be defect free, and magnetic moments were shown to be absent in fully relativistic \textit{ab initio} band structure calculation, leading to unclear situation.\cite{dey_PRB_2016} However, in further studies the weak response was attributed to a small fraction of paramagnetic centres ($\approx 4$\%), primarily due to Ir$^{6+}$ ($S = 3/2$) defects that exhibit magnetic interactions below 20 K.\cite{fuchs_PRL_2018}

Resonant inelastic X-ray scattering experiments on momentum dependence of spin-orbital excitations in both $\rm Sr_2YIrO_6$ and $\rm Ba_2YIrO_6$ showed excitations at about 370 and 650 meV.\cite{kusch_PRB_2018} These excitations were suggested to correspond to the triplet ($J = 1$) and quintet ($J = 2$) excitations, respectively. 

Recently, the cubic antifluorite type halides $\rm K_2OsCl_6$ and $\rm K_2OsBr_6$ have emerged as another group of $5d^4$ compounds which exhibit intra-$t_{2g}$ excitations at approximately 340 meV and 600 meV energy in both compounds,\cite{warzanowski_PRB_2023} which are similar to the excitation energies observed in the double perovskite compounds. No evidence of magnetic defects was found in these compounds, and the estimated magnetic susceptibility using second-order perturbation theory in the $J=0$ state was found to be in good agreement with the measured temperature independent susceptibility $\approx 10 \times 10^{-4}$ emu/mol.

The non-magnetic $J=0$ ground state in strongly spin-orbit coupled systems has been investigated in earlier theoretical works.\cite{khaliullin_PRL_2013,nag_PRB_2018,hoshin_JMMM_2018,zhang_PRB_2022} Excitonic condensation was suggested as the cause of unconventional magnetism rather than preexisting local moments in the $4d$ and $5d$ TM ions with $t_{2g}^4$ configuration.\cite{khaliullin_PRL_2013} By fitting the measured RIXS spectrum for $\rm Ba_2YIrO_6$ with an atomic model, an estimated value $\lambda \approx 0.4$ eV was obtained for the bare SOC.\cite{nag_PRB_2018} Non-magnetic ground state without defects or anti-site disorder was reported for multi-site clusters of $\rm Ba_2YIrO_6$ using exact diagonalization.\cite{hoshin_JMMM_2018} The $J=0$ state in the cubic halide compounds $\rm K_2OsX_6$ was recently predicted in density functional theory investigation.\cite{zhang_PRB_2022}

A systematic evolution with increasing SOC from the AFM order to the non-magnetic $J=0$ state, showing clear transition at the critical SOC value has so far not been investigated. In the following, we present results of the generalised self-consistent + fluctuations approach using the realistic three-orbital interacting electron model.

\begin{figure}[t]
\vspace*{0mm}
\hspace*{0mm}
\psfig{figure=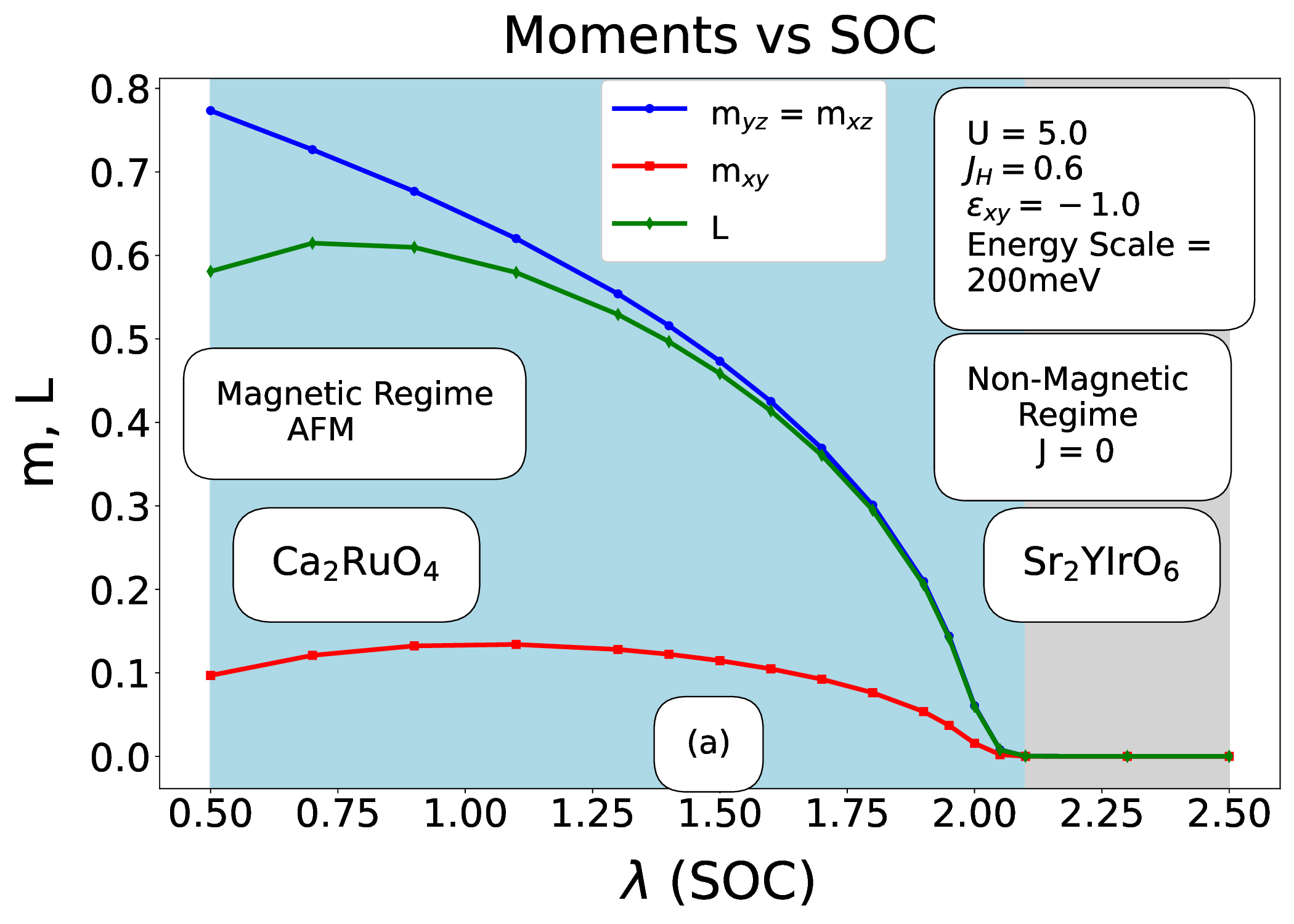,angle=0,width=80 mm}
\psfig{figure=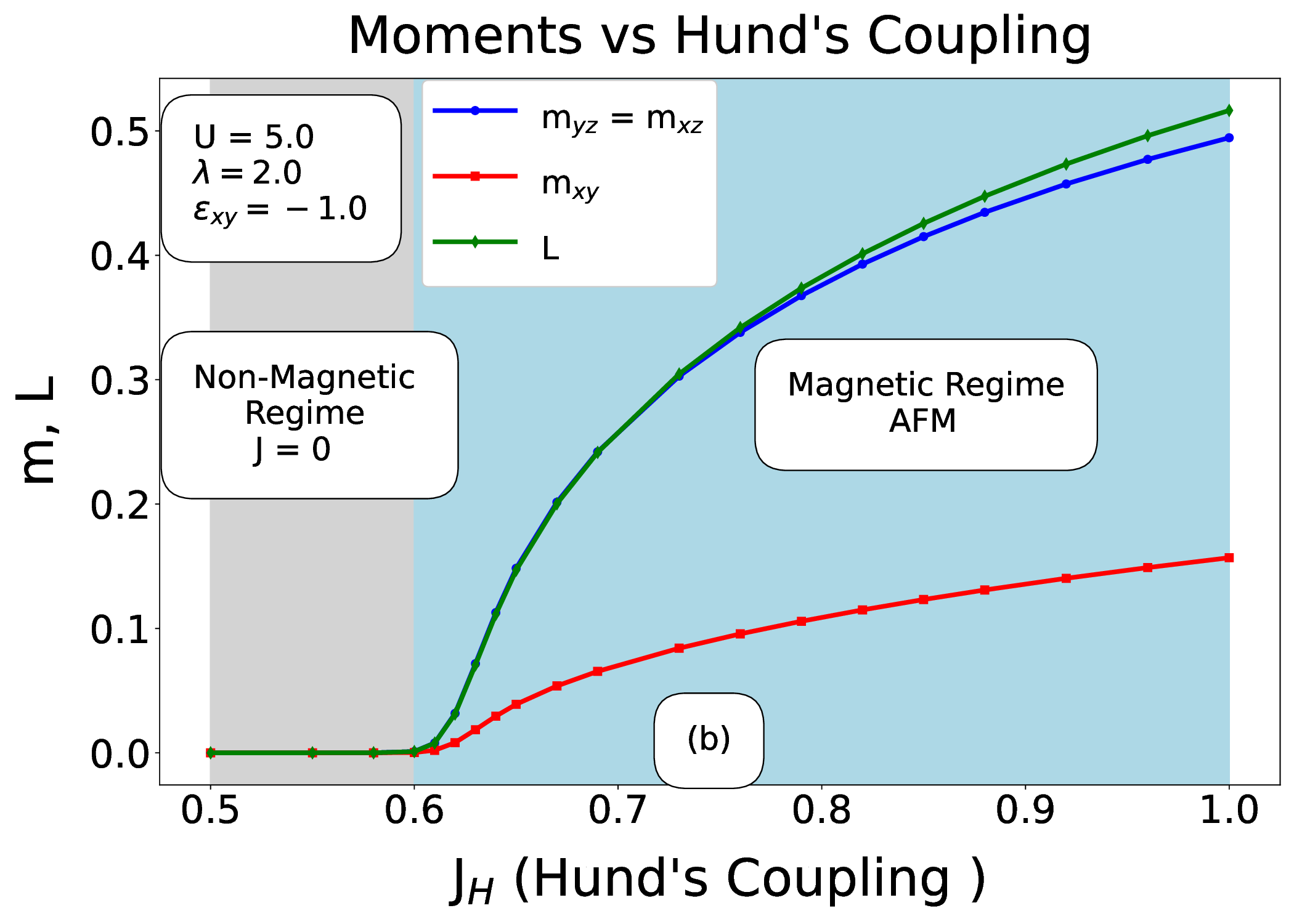,angle=0,width=80 mm}
\caption{(a) With increasing SOC strength, the spin and orbital moments continuously decrease to zero due to enhanced spin-orbital entanglement, showing the onset of the $J=0$ state at $\lambda=\lambda^*$. (b) Similar behaviour but with decreasing Hund's coupling reflecting the antagonistic effect of $J_{\rm H}$ on spin-orbital entanglement.} 
\label{fig2}
\end{figure} 

The behaviour of calculated spin and orbital moments with increasing SOC is shown in Fig. \ref{fig2}(a). The spin and orbital moments presented here correspond to the magnitudes $m_\mu =[\sum_{\alpha=x,y,z}(m_\mu ^\alpha)^2]^{1/2}$ for the three orbitals $\mu=d_{yz},\;d_{xz},\;d_{xy}$ and $L=[\sum_{\alpha} L_\alpha^2]^{1/2}$. Both the spin and orbital moments decrease (a) with SOC reflecting the increasing spin-orbital entanglement, and vanish at a critical SOC strength $\lambda=\lambda^*$ signalling the onset of the $J=0$ state. For fixed SOC, similar behaviour is seen Fig. \ref{fig2}(b) with decreasing Hund's coupling due to the antagonistic effect of $J_{\rm H}$ on spin-orbital entanglement. The above result provides continuous interpolation from AFM order to the $J=0$ state. For energy scale 200 meV, the corresponding $U$ value is 1 eV, $J_{\rm H}\approx U/8$ as taken in recent studies of other $5d$ systems,\cite{mohapatra_JPCM_2021} and the critical SOC value $\lambda^*=400$ meV. The value $\epsilon_{xy}=-1$ was taken here to allow for proper interpolation to AFM order for weaker SOC corresponding to $\rm Ca_2RuO_4$. The critical SOC value was found to be nearly independent of the crystal field in a wide range $0<|\epsilon_{xy}|<3$. 

\begin{figure}[t]
\vspace*{0mm}
\hspace*{0mm}
\psfig{figure=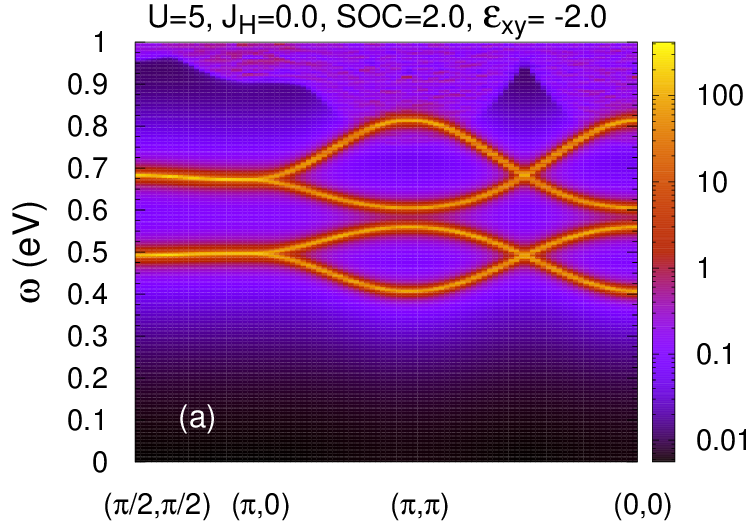,angle=0,width=80 mm}
\psfig{figure=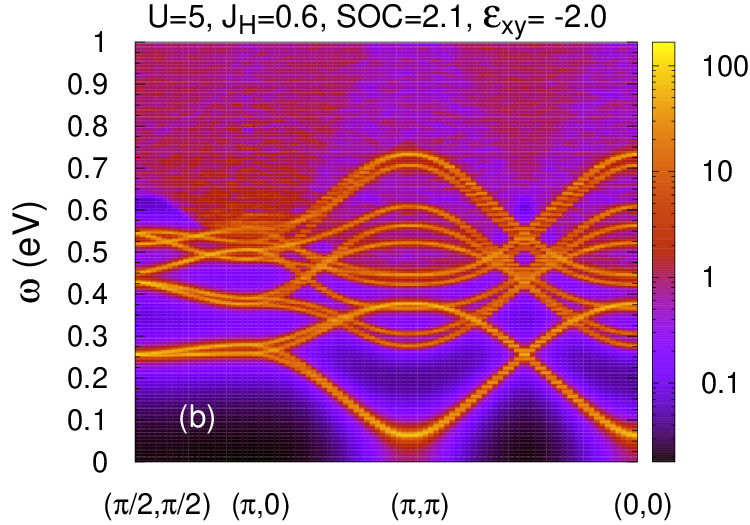,angle=0,width=80 mm}
\caption{(a) The spectral function of collective excitations deep in the $J=0$ state shows well defined spin-orbit exciton modes corresponding to particle-hole excitations between the SOC-induced $j=3/2,1/2$ sector states. (b) Just into the $J=0$ state ($\lambda \gtrsim \lambda^*$), the extremely low energy of the lower exciton mode at wave-vector ${\bf q}=(0,0)$ or $(\pi,\pi)$ will go to zero at $\lambda = \lambda^*$ and become negative for $\lambda < \lambda^*$, signalling instability of the $J=0$ state. The energy scale $|t_1|=200$ meV.} 
\label{fig3}
\end{figure} 

The electronic band structure in the $J=0$ state shows SOC-induced spin-orbital entangled states with the four $j=3/2$ sector states occupied and the two $j=1/2$ sector states unoccupied. The spectral function of collective excitations shows well defined spin-orbit exciton modes as seen in Fig. \ref{fig3}(a), where $J_{\rm H}=0$ for simplicity, and we have taken $\epsilon_{xy}=-2$ in order to enhance the energy separation between the exciton modes for clarity. These exciton modes correspond to particle-hole excitations across the Coulomb renormalised spin-orbit gap between the $m_j=\pm 3/2,\pm 1/2$ (hole) states of the $j=3/2$ sector and the $m_j=\pm 1/2$ (particle) states of the $j=1/2$ sector.\cite{mohapatra_JMMM_2020} Repeated attractive interaction between the propagating particle-hole pair in the resonant scattering process lowers the excitation energy from the renormalised spin-orbit gap depending on the $(j,m_j)$ states.\cite{mohapatra_JMMM_2020} Analysis of the fluctuation propagator eigenvector in the orbital-pair basis reveals that the dominant contribution involves $d_{yz}/d_{xz}$ orbitals in the lower energy mode and $d_{xy}/d_{yz}$ and $d_{xy}/d_{xz}$ orbitals in the higher energy mode. 

The multiplet structure of the excitons is clearly revealed at finite Hund's coupling $J_{\rm H}$ as seen in Fig. \ref{fig3}(b). The eight possibilities for particle-hole excitations involving hole in the four $j=3/2$ states and particle in the two $j=1/2$ states accounts for the total eight-fold multiplicity of the $J=1$ (triplet) and $J=2$ (quintet) excitations. Due to Brillouin zone folding in the two sublattice-basis used, momenta ${\bf q}=(0,0)$ and $(\pi,\pi)$ are equivalent, resulting in an extra multiplicity of two.
 
\begin{figure}[t]
\vspace*{0mm}
\hspace*{0mm}
\psfig{figure=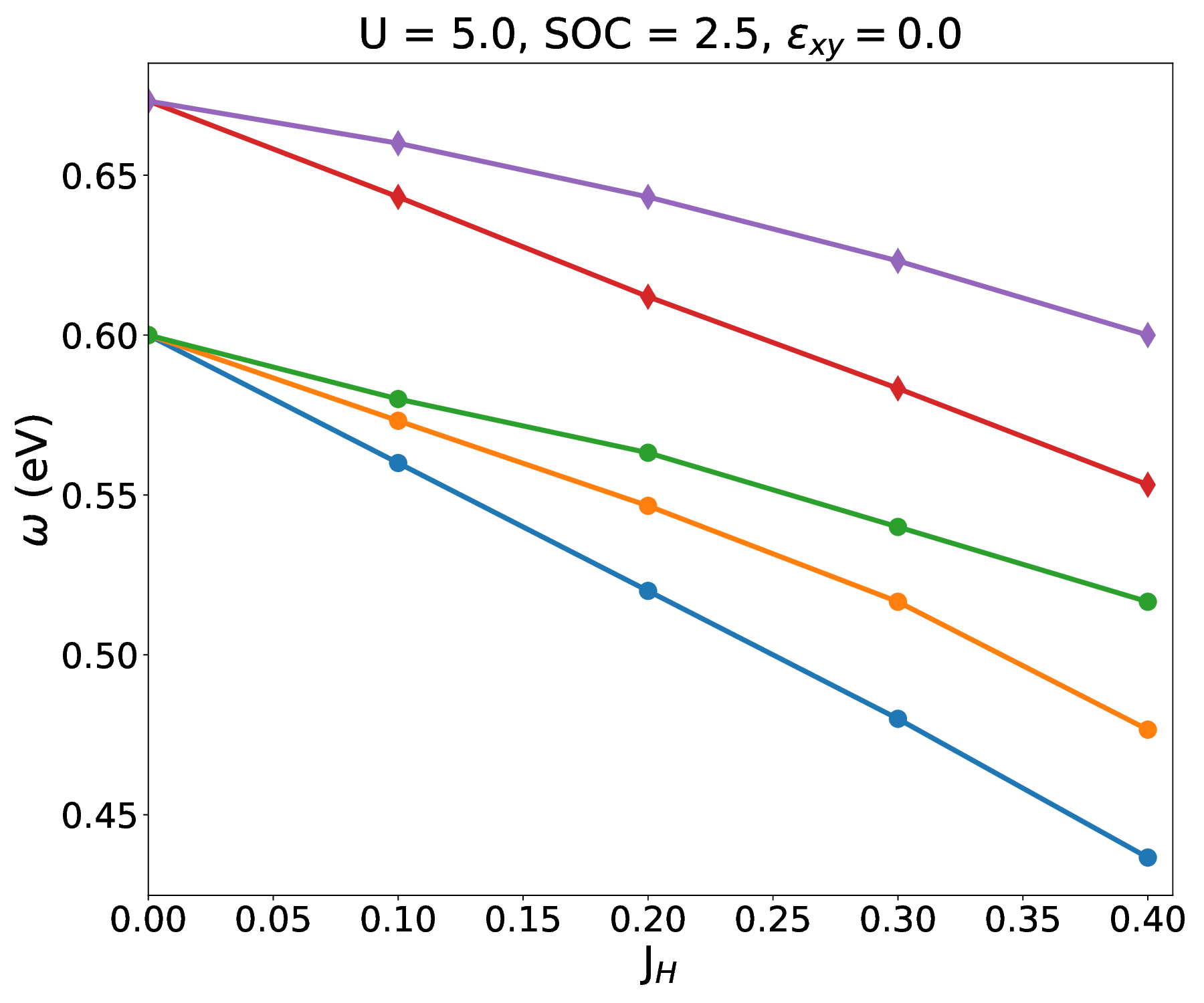,angle=0,width=80 mm}
\psfig{figure=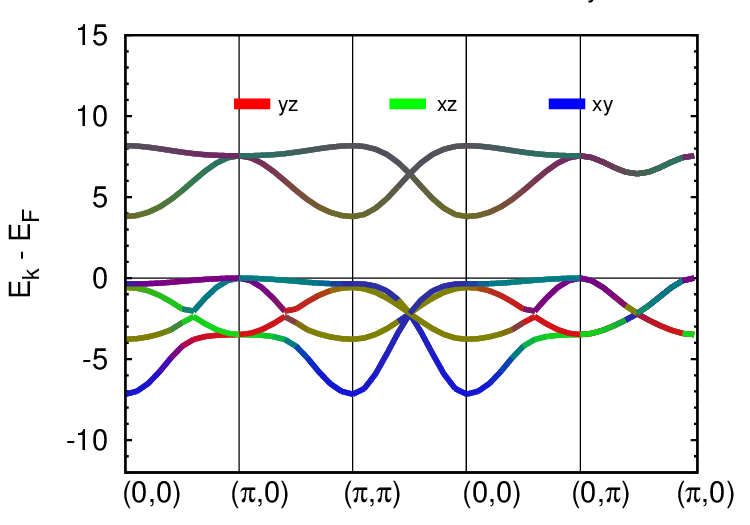,angle=0,width=80 mm}
\caption{(a) Behaviour of exciton energies at ${\bf q}=(\pi/2,\pi/2)$ with Hund's coupling clearly shows two groups of excitations, corresponding to $J=2$ modes (lower, $m_J=\pm 2,\pm 1,0$) and $J=1$ modes (upper, $m_J=\pm 1,0$). (b) Orbital-resolved electronic band structure in the $J=0$ state for $J_{\rm H}=0$.} 
\label{fig3plus}
\end{figure}  
 


We have also calculated the exciton energies for realistic SOC value $\lambda=500$ meV and $\epsilon_{xy}=0$ (no energy offset for $d_{xy}$ orbital). At fixed wave vector ${\bf q}=(\pi/2,\pi/2)$, the behaviour of exciton energies with Hund's coupling is shown in Fig. \ref{fig3plus}(a), which clearly allows for identification of the three lower-energy excitons and two higher-energy excitons as corresponding to $J=2$ modes $(m_J=\pm 2,\pm 1,0)$ and $J=1$ modes $(m_J=\pm 1,0)$, respectively. For realistic $J_{\rm H}=0.6$ as in Fig. \ref{fig3}(b), the excitation energies lie in the range $\sim$ 400 to 600 meV, which is in good agreement with the measured excitation energies $\sim 350$ and $\sim 600$ meV in RIXS experiments. 

An estimate of the exciton energy based on the picture mentioned above (repeated attractive interaction between the propagating particle-hole pair excited across the Coulomb renormalised spin-orbit gap) yields: $\omega_{\rm exc}=\Delta E_{\rm SOC}-U\approx$ (1.6-1) eV=0.6 eV, which agrees well with the calculated result as seen in Fig. \ref{fig3plus}(a) for $J_{\rm H}=0$. Here, for $\Delta E_{\rm SOC}$ we have taken the energy difference between the $j=3/2$ and $j=1/2$ sector band energies at ${\bf k}=(\pi/2,\pi/2)$ which is approximately $8\times 200$ meV = 1.60 eV from Fig. \ref{fig3plus}(b) showing the electronic band structure in the $J=0$ state. 


\section{Instability of the $J=0$ state}

The $J=0$ state is stabilised for SOC greater than a critical value $(\lambda > \lambda^*)$ for fixed Hamiltonian parameters. Fig. \ref{fig3}(b) shows the spectral function of collective excitations in the $J=0$ state for SOC just above the critical value ($\lambda \gtrsim \lambda^*$). As seen here, the excitation energy of the lower exciton branch is nearly zero at wave-vector ${\bf q}=(0,0)$/$(\pi,\pi)$. 

With $\lambda$ approaching $\lambda^*$ from above, this excitation energy drops to zero at $\lambda = \lambda^*$, and will extrapolate to negative values for $\lambda < \lambda^*$, signalling instability of the $J=0$ state. Since spin and orbital moments also spontaneously appear at $\lambda < \lambda^*$, the instability is therefore towards local moment formation. Thus, Bose condensation in the negative-energy exciton modes for $\lambda < \lambda^*$ is equivalent to local moment formation. 

For $\lambda \gtrsim \lambda^*$, the system is in the $J=0$ state but on the verge of local-moment formation. Due to the low exciton energy in this case, significant thermal excitation of these modes at finite temperature would result in emergence of fluctuating local moments, leading to enhanced magnetic susceptibility. This temperature dependent Van Vleck type behaviour is expected in the $\lambda \gtrsim \lambda^*$ regime which does not support magnetic moments. 


\section{Half-metallic FM order: Application to $\rm 3d^4$ and $\rm 4d^4$ systems}

As mentioned earlier, half-metallic ground state has been reported for both the cubic $3d^4$ compound $\rm BaFeO_3$\cite{hoedl_JPCC_2022} and the layered $4d^4$ compound $\rm Sr_2RuO_4$.\cite{huang_SCIREP_2020} While strong FM order accompanies the half-metallic state in $\rm BaFeO_3$, it was suggested that in $\rm Sr_2RuO_4$ the FM order is unstable due to spin fluctuations in view of extremely close energy of the antiferromagnetic phase. In the following, we will apply our generalized self-consistent + fluctuations approach for a comparative study of these two compounds, focussing on the high-energy spin-orbital excitations in $\rm Sr_2RuO_4$. 

Initial RIXS experiments on $\rm Sr_2RuO_4$ using combined X-ray absorption and oxygen K-edge scattering showed excitations near 350 meV.\cite{fatuzzo_PRB_2015} Recent low temperature (25 K) RIXS studies in the Fermi liquid regime using Ru $L_3$-edge resonant scattering show complex spin and orbital excitations.\cite{suzuki_NATCOM_2023} Low-energy spin excitations confined below $\sim 200$ meV were identified along with high-energy orbital fluctuations with intensity peaking near $500$ meV but extending to higher energies up to $\sim 1$ eV. 

Very recent laser angle resolved photo emission spectroscopy (ARPES) experiments and density functional theory (DFT) calculations on $\rm Sr_2RuO_4$ have highlighted the significant role of SOC on the electronic band and Fermi surface structure.\cite{kondo_PRB_2024,tamai_PRX_2019,kim_PRL_2018} Compared to the bare SOC value taken ($\sim 100$ meV), the effective SOC was found to be significantly enhanced due to Coulomb interaction, and the measured Fermi surface structure was found to be well reproduced with the enhanced SOC value ($\sim 200$ meV) for zero crystal field splitting. The Coulomb interaction values considered in these studies are: $U\sim 2$ eV, $J_{\rm H}/U \sim 1/5$.




\begin{figure}[t]
\vspace*{0mm}
\hspace*{0mm}
\psfig{figure=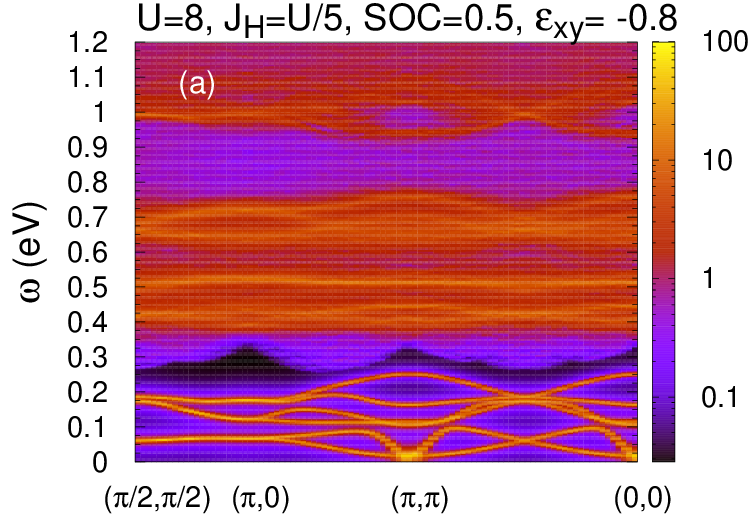,angle=0,width=53 mm}
\psfig{figure=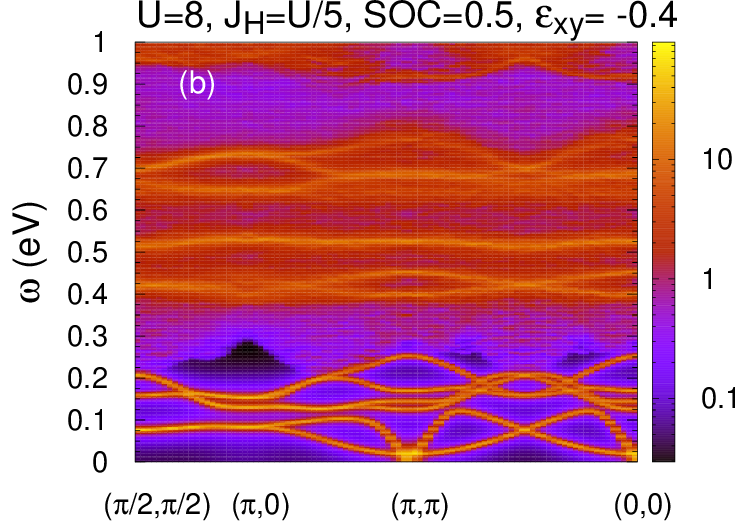,angle=0,width=53 mm}
\psfig{figure=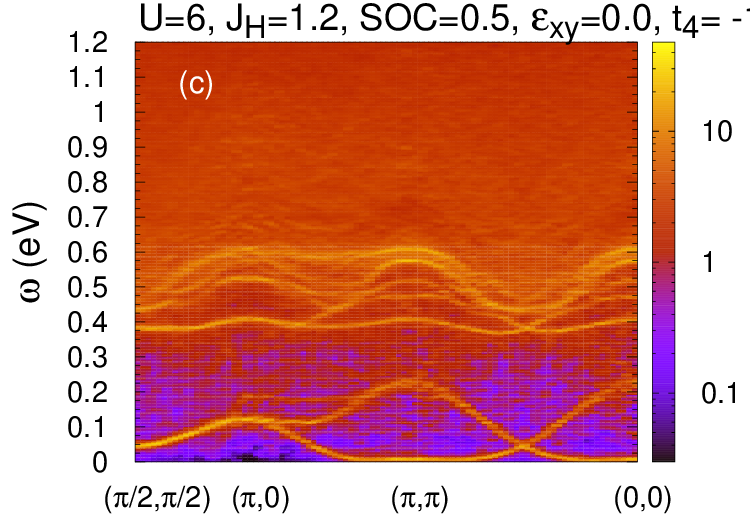,angle=0,width=53 mm}
\caption{Evolution of the spectral function of collective excitations in the $4d^4$ system (intermediate SOC strength) with decreasing crystal field $\epsilon_{xy}$: (a) strong, (b) intermediate, (c) zero. The high-energy excitations near 500 meV in case (c) corresponding to $\rm Sr_2RuO_4$ are seen to evolve from spin-orbitons and spin-orbit excitons in the case (a) corresponding to $\rm Ca_2RuO_4$.} 
\label{fig4}
\end{figure} 


Fig. \ref{fig4} shows the evolution of the calculated spectral function of collective excitations with decreasing crystal field: (a) strong ($\epsilon_{xy}=-0.8$), (b) intermediate ($\epsilon_{xy}=-0.4$), and (c) zero ($\epsilon_{xy}=0$). The two end cases (a) and (c) correspond to $\rm Ca_2RuO_4$ (AFM order) and $\rm Sr_2RuO_4$ (half metallic FM order), respectively. The hopping energy scales considered are 200 meV for (a,b) and 250 meV for (c) to account for the lattice compression in the $\rm Sr_2RuO_4$ crystal due to larger size of Sr ion. In increasing order of energy, the collective excitation modes in case (a) correspond to magnon, two orbitons, two spin-orbitons, spin-orbit excitons, and (near 1 eV) the out-of-phase magnon due to Hund's coupling. 

In the half-metallic case as relevant for $\rm Sr_2RuO_4$, there is a single low-energy magnon mode as seen in Fig. \ref{fig4}(c) with excitation energy up to $\sim$ 200 meV for zero crystal field ($\epsilon_{xy}=0$). This low-energy magnon mode represents in-phase transverse spin fluctuations as all spin moments $m_\mu$ ($\mu=d_{yz},d_{xz},d_{xy}$) are coupled due to the Hund's coupling term. Strong high-energy excitations in the range 400-600 meV are also present, which are seen to evolve from the spin-orbitons and spin-orbit excitons in the case (a) corresponding to $\rm Ca_2RuO_4$. The calculated excitation energies of both spin and orbital excitations and their energy separation are in good agreement with recent RIXS measurements in $\rm Sr_2RuO_4$,\cite{suzuki_NATCOM_2023} and with the characterisation of $\rm Sr_2RuO_4$ as a Hund metal.\cite{blesio_PRR_2024}

Fig. \ref{fig4}(c) also shows instability of long-range FM order. Instead of the characteristic $\omega_{\bf q}=Dq^2$ type magnon energy behaviour for small $q$ ($D$ refers to spin stiffness), the calculated magnon energy is seen to go to zero at finite wave vector ${\bf q}\sim(\pi/3,\pi/3)$ and extrapolates to negative energy near $(0,0)$. Negative magnon energy for small ${\bf q}$ signals instability of long-range FM order with respect to long-wavelength spin twisting which will result in energy lowering. Note that the magnon dispersion shows the characteristic Brillouin zone folding due to the two-sublattice basis used in the calculation. 

The above behaviour of magnon excitations is broadly consistent with inelastic neutron scattering studies on $\rm Sr_2RuO_4$. Besides incommensurate spin fluctuations due to nesting in the quasi-one dimensional $d_{yz},d_{xz}$ bands,\cite{jenni_PRB_2021} magnetic fluctuations centred at the Brillouin zone origin which are typically associated with ferromagnetism have also been reported.\cite{steffens_PRL_2019} However, as our focus is on the high-energy nearly local spin-orbital excitations, detailed investigation into these low-energy features is beyond the scope of this paper. 

\begin{figure}[t]
\vspace*{0mm}
\hspace*{0mm}
\psfig{figure=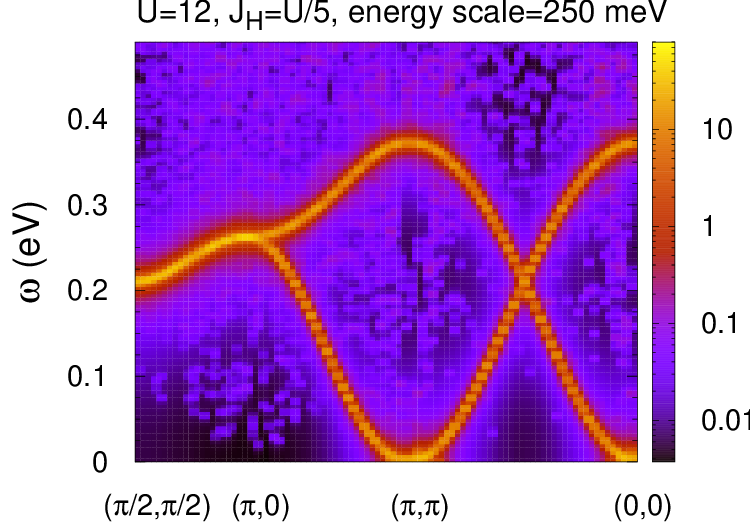,angle=0,width=75 mm,angle=0}
\caption{Magnon dispersion in the semi-metallic state with FM order for the one-orbital model ($d_{xy}$) at 4/3 electron filling, showing stable ferromagnetic state as evidenced by positive magnon energies over entire Brillouin zone. A cubic compound such as $\rm BaFeO_3$ with degenerate $d_{xy},d_{yz},d_{xz}$ orbitals would yield qualitatively similar magnon dispersion.} 
\label{fig5}
\end{figure}

The instability of FM order in $\rm Sr_2RuO_4$ partly arises from reduced $U/|t_1|$ in this $4d$ system. Ferromagnetism is a strong coupling phenomenon favoured by large $U$ as in $3d$ systems. Another factor is the frustration of FM order by competing magnetic interactions generated by the quasi-one dimensional $d_{yz},d_{xz}$ orbitals. Indeed, when the $d_{yz}$ and $d_{xz}$ orbitals are rendered ineffective (by pushing up their energy to infinity), the calculated magnon energies are positive over the entire Brillouin zone, as seen in Fig. \ref{fig5}. A qualitatively similar magnon dispersion would be obtained for a cubic compound such as $\rm BaFeO_3$ with degenerate $d_{yz},d_{xz},d_{xy}$ orbitals and identical hopping terms for all three orbitals, as explained below. As appropriate for the $3d$ system, a stronger interaction term $U/|t_1|=12$ has been considered in Fig. \ref{fig5}, which also favours stable FM order. 

In the above cubic system, the emergent nearest-neighbour (NN) FM spin couplings in the $x-y$ plane will be generated by itinerant electron dynamics in the $d_{xy}$ orbital, and similarly in the other two planes by electrons in the degenerate $d_{yz}$ and $d_{xz}$ orbitals. Magnon dispersion will thus be identical in the three momentum planes $q_x - q_y$, $q_y - q_z$, $q_z - q_x$, and effective spin couplings will be simply doubled due to contribution from all three orbitals. For the cubic compound $\rm BaFeO_3$, the magnon energies will therefore be positive over the entire Brillouin zone, confirming that long-range FM order is strongly stabilised. 

This provides fundamental insight into the stability of long-range FM order in the cubic compound $\rm BaFeO_3$, while competing interactions due to quasi-one dimensional electron hopping in $d_{yz},d_{xz}$ orbitals in the layered $\rm Sr_2RuO_4$ compound renders the FM order marginally unstable and consequent proclivity towards incommensurate order. Thus, while $\rm BaFeO_3$ is a strong ferromagnet with high Curie temperature, $\rm Sr_2RuO_4$ is characterised as a Fermi liquid (with low Fermi temperature due to low spin excitation energies). The finding of commensurate FM order in weakly Fe-doped $\rm Sr_2RuO_4$ demonstrates that the undoped compound is on the verge of exhibiting stable FM order.\cite{zhu_PRB_2017}



\begin{figure}[t]
\vspace*{-10mm}
\hspace*{0mm}
\psfig{figure=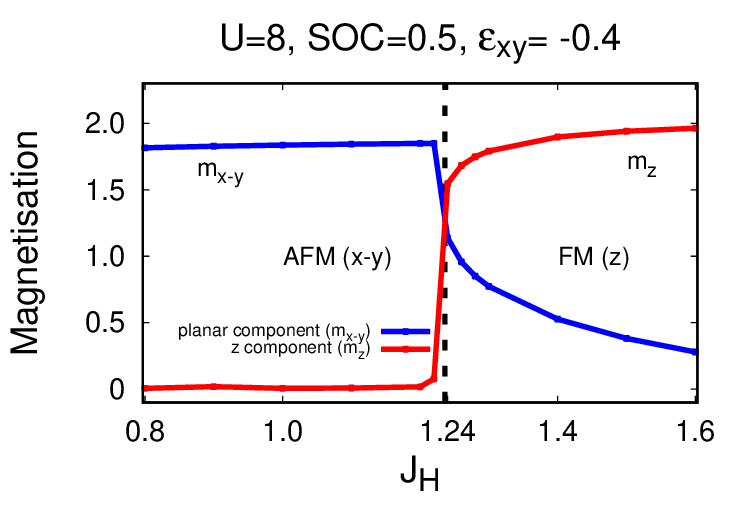,angle=0,width=80mm}
\vspace{-10mm}
\caption{Variation of magnetisation components (planar and $z$) with Hund's coupling, showing transition from FM ($z$) order to AFM (planar) order with decreasing $J_{\rm H}$.} 
\label{fig6}
\end{figure} 

The significant role of Hund's coupling on magnetic order is illustrated in Fig. \ref{fig6} which shows transition from FM($z$) order to AFM (planar) order with decreasing $J_{\rm H}$. This transition is due to enhanced Coulomb renormalised tetragonal splitting for $J_{\rm H} < U/5$,\cite{mohapatra_JPCM_2020}
\begin{equation}
\tilde{\epsilon}_{yz,xz} - \tilde{\epsilon}_{xy} = \epsilon_{yz,xz} - \epsilon_{xy} + \frac{U-5J_{\rm H}}{2} \langle n_{xy} - n_{yz,xz} \rangle
\end{equation}
resulting in enhanced electron density in $d_{xy}$ orbital ($n_{xy}\rightarrow 2$) and decreasing $d_{yz,xz}$ orbital densities ($n_{yz,xz}\rightarrow 1$), thus favouring the AFM (planar) order with decreasing $J_{\rm H}$. 

On the other hand, with increasing $J_{\rm H}$, the Coulomb contribution to tetragonal splitting vanishes when $J_{\rm H}=U/5$. The electron densities therefore become nearly identical $n_\mu \approx 4/3$ in all three orbitals for negligible bare tetragonal splitting, leading to emergence of FM spin couplings. This highlights the significant role of $J_{\rm H}$ in stabilising the half-metallic state (with short-range FM order) in $\rm Sr_2RuO_4$.

\begin{figure}[t]
\vspace*{-10mm}
\hspace*{0mm}
\psfig{figure=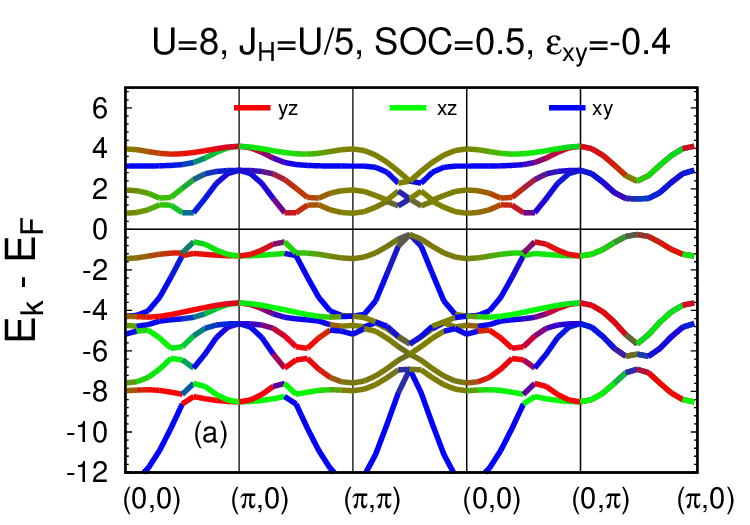,angle=0,width=53mm,angle=0}
\psfig{figure=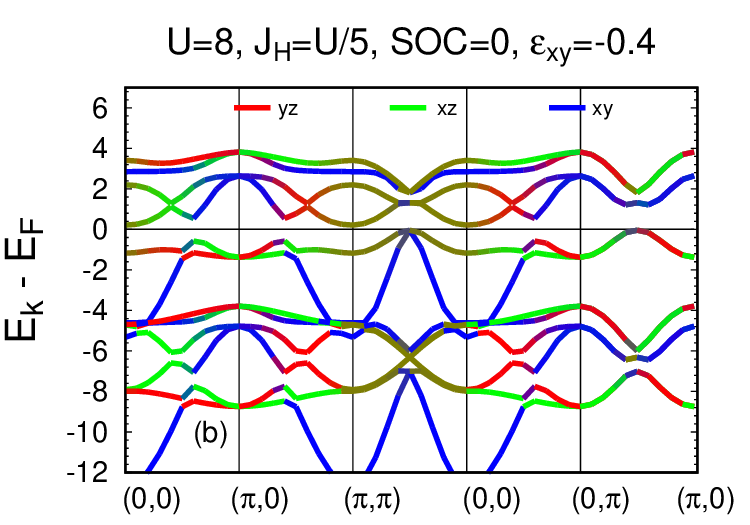,angle=0,width=53mm,angle=0}
\psfig{figure=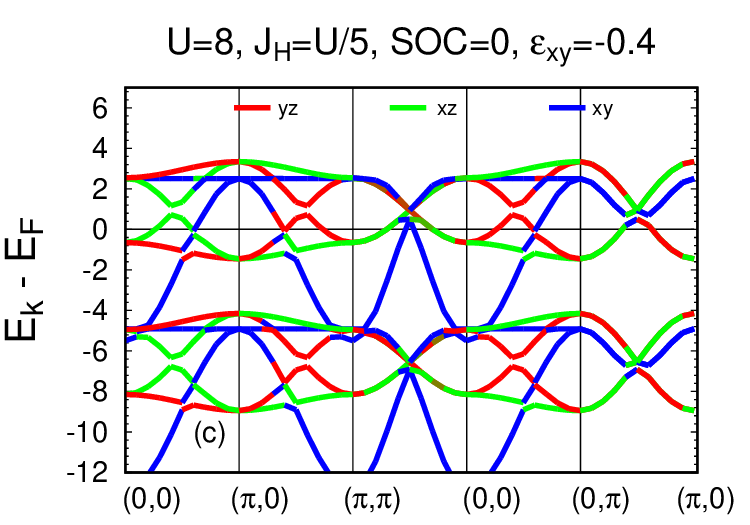,angle=0,width=53mm,angle=0}
\caption{Orbital resolved electronic band structure in the self-consistent state with: (a) finite bare SOC, (b) bare SOC set to zero, and (c) orbital mixing condensates not included. Near the Fermi energy, while band splitting near $(\pi/2,\pi/2)$ between entangled states involving $d_{yz},d_{xz}$ orbitals remains finite even for zero bare SOC (b) due to Coulomb interaction induced SOC terms, the splitting vanishes when the orbital mixing condensates are not included.} 
\label{fig7}
\end{figure}

Turning now to the role of SOC, we have investigated the electronic band structure for the intermediate case $\epsilon_{xy}=-0.4$ with and without SOC, and also when the orbital mixing condensates are not included. These results are shown in Fig. \ref{fig7}. Focussing on the band structure near the Fermi energy, the small splitting seen near $(\pi/2,\pi/2)$ between the entangled states involving $d_{yz},d_{xz}$ orbitals is seen to decrease when bare SOC is reduced to zero (a)$\rightarrow$(b). However, the splitting is finite even for zero bare SOC (b), and is seen to identically vanish only when the orbital mixing condensates are not included (c). 

This evolution of the small splitting confirms the role of the Coulomb renormalised SOC component $\lambda_z$. Due to the spontaneous generation of Coulomb interaction induced SOC-like term $\lambda_z^{\rm int}$ (see Eq. \ref{phys_quan}), this splitting is present even in the absence of the bare SOC term. These results highlight the critical role of SOC in the electronic band structure near the Fermi energy. 

\begin{figure}[t]
\vspace*{0mm}
\hspace*{0mm}
\psfig{figure=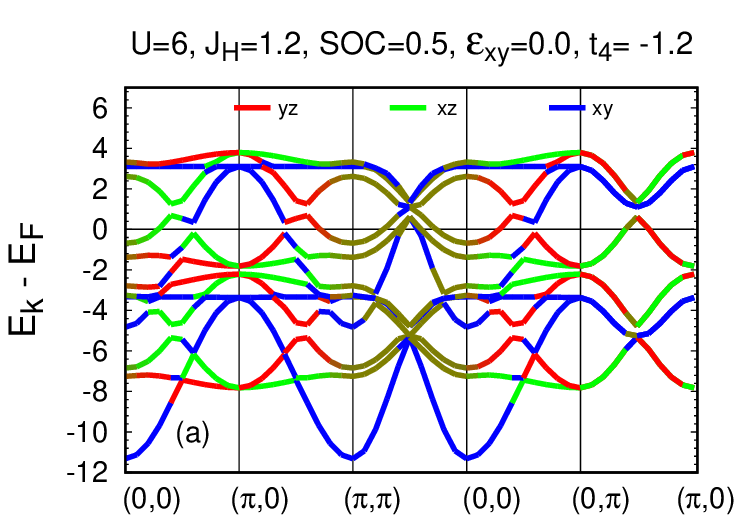,angle=0,width=70mm}
\psfig{figure=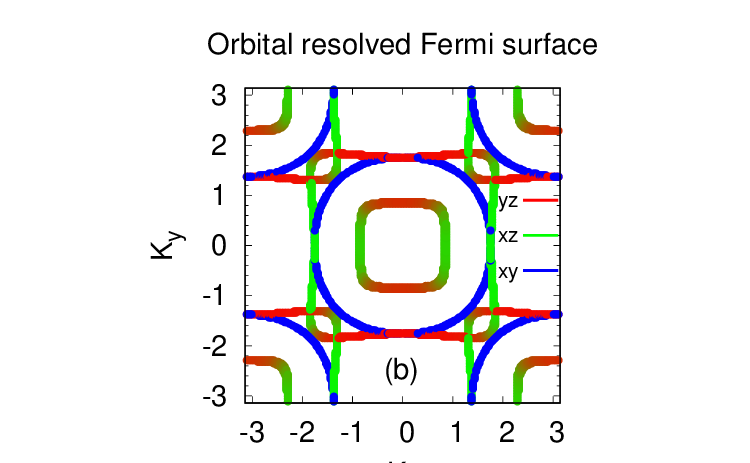,angle=0,width=90mm}
\caption{Calculated (a) electronic band structure and (b) Fermi surface structure in the self-consistent half-metallic state with FM order for the parameter set corresponding to $\rm Sr_2RuO_4$.} 
\label{fig8}
\end{figure}

The calculated orbital resolved electronic band structure and the Fermi surface structure in the self-consistent half-metallic state with FM order are shown in Fig. \ref{fig8} for the parameter set corresponding to $\rm Sr_2RuO_4$. It should be noted that parts within the reduced Brillouin zone get mapped to those lying outside due to Brillouin zone folding in the two-sublattice basis used. Near the Fermi energy, there are two quasi-one dimensional ($\alpha,\beta$) bands involving $d_{yz},d_{xz}$ orbitals, and one two-dimensional ($\gamma$) band involving $d_{xy}$ orbital. These electronic band structure features are in good agreement with quantum oscillation and ARPES measurements.\cite{tamai_PRX_2019,mackenzie_PRL_1996,damascelli_PRL_2000} We find that for bare SOC $\lambda=0.5$, the renormalised SOC component $\lambda_z \approx 0.8=200$ meV (using energy scale 250 meV), which is in excellent agreement with the enhanced SOC value 200 meV considered in the fitting with ARPES data.\cite{tamai_PRX_2019} 


The role of van Hove singularity on the stability of FM order in a metallic ferromagnet has been investigated earlier in the context of quantum corrections to the magnon propagator.\cite{pandey_PRB_2007} For a one-band model with electronic band dispersion ${\cal E}_{\bf k}$, the momentum gradient components $\nabla_{\bf k} {\cal E}_{\bf k}$ are small over an extended ${\bf k}$ region near the saddle points, which significantly suppresses the quantum correction to spin stiffness arising from self-energy and vertex corrections, leading to stability of FM order. Quantum corrections are also strongly suppressed in a ${\cal N}$ orbital model, where the suppression factor is $1/{\cal N}$ in the orbitally symmetric limit $J_{\rm H}=U$.\cite{pandey_PRB_2008} It should be noted that for a one-band model the hopping signs are reversed and the band filling transforms from $n=4/3$ to $n=2/3$ under the particle-hole transformation. 

\section{Discussion and Conclusions}

Using a common theoretical framework for the $3d^4$, $4d^4$, and $5d^4$ compounds, the unified view of the evolution of spin-orbital order and collective excitations in different parameter regimes corresponding to the different compounds provides novel insights into the interplay of SOC, crystal field, and Coulomb interaction. 

For example, in the intermediate state ($\epsilon_{xy}=-0.4$) between $\rm Ca_2RuO_4$ and $\rm Sr_2RuO_4$, there is an antiferro ordering of planar orbital moments $\langle L_x\rangle$ and $\langle L_y\rangle$, leading to a complex excitation spectrum of orbitons and magnon in the same low-energy range (Fig. \ref{fig4}b). In a striking transformation, this complexity disappears in the half-metallic state at $\epsilon_{xy}=0$ where the planar orbital moments identically vanish $\langle L_x\rangle=\langle L_y\rangle=0$, leaving a single ferromagnetic magnon excitation. Since there are no low-energy orbitons, the Hund's metal is realised with characteristic large energy difference between the low-energy spin excitations and high-energy orbital excitations (spin-orbitons and spin-orbit excitons). 

The nominally $S=1$ AFM order in $\rm Ca_2RuO_4$ (strong crystal field, intermediate SOC and Hubbard $U$) evolves to robust half-metallic FM order in $\rm BaFeO_3$ (no crystal field, weak SOC, strong $U$) and half-metallic marginally unstable FM order in $\rm Sr_2RuO_4$ (no crystal field, intermediate SOC and $U$), where the instability arises from competing spin interactions generated by electron dynamics in the $d_{yz},d_{xz}$ orbitals. The calculated spectral function of collective excitations shows low-energy magnon excitations up to 200 meV and high-energy excitations near 500 meV. The calculated excitation energies are in good agreement with recent RIXS measurements in $\rm Sr_2RuO_4$.

In the half-metallic state of $\rm Sr_2RuO_4$, orbital mixing terms involving $d_{xy}$ orbital were found to be identically zero, leading to vanishing orbital moments $\langle L_x\rangle=\langle L_y\rangle=0$. On the other hand, strong entanglement between $d_{yz},d_{xz}$ orbitals resulted in splitting of electronic bands near Fermi energy due to strong Coulomb interaction induced enhancement of SOC, leading to robust features in the band structure and Fermi surface structure. Indeed, the band splitting was found to be present even in the absence of bare SOC due to spontaneous generation of SOC-like terms from Coulomb interaction, further underlining the complex interplay of electronic correlations with spin and orbital degrees of freedom in $\rm Sr_2RuO_4$.
 
Similarly, the evolution with increasing SOC leading to the onset of the $J=0$ state reveals novel features such as the multiplet structure of the $J=1$ and $J=2$ exciton modes at finite $J_{\rm H}$. Our study of collective spin-orbital excitations further showed that the lowest spin-orbit exciton mode energy approaches zero when $\lambda=\lambda^*$ and increases sharply for $\lambda>\lambda^*$, whereas the local spin and orbital moments increase sharply for $\lambda<\lambda^*$ where the extrapolated exciton mode energy turns negative. Similar continuous interpolation from finite-moment to zero-moment regime was seen with decreasing Hund's coupling $J_{\rm H}$, thus providing deeper understanding of the evolution of spin-orbital order and excitations. 

The calculated critical value $\lambda^*\approx 400$ meV of the bare SOC lies in the realistic SOC range for the $5d$ ions (Ir, Os) indicating that $\lambda>\lambda^*$ is the appropriate regime for the $5d^4$ double perovskite and cubic halide compounds in which the $J=0$ ground state with no spin or orbital moments is stabilised  due to strong SOC and weak Coulomb interaction. Collective excitations in the $J=0$ state yield well defined propagating modes corresponding to the spin-orbit exciton modes involving particle-hole excitations across the Coulomb renormalised spin-orbit gap between the $j=3/2$ and $j=1/2$ sector states. The calculated excitation energies are in good agreement with recent RIXS measurements ($\sim 350$ and 600 meV) in $\rm Sr_2YIrO_6$ and $\rm K_2OsCl_6$).

In the $\lambda\gtrsim \lambda^*$ regime, temperature-dependent Van Vleck type behaviour arising from thermal excitation of the low-energy exciton mode at finite temperature resulting in qualitative change in the magnetic response should be of significant interest. Also, in analogy with the very recent photo-excitation studies showing magnetic transition induced by transiently reduced crystal field in $\rm Ca_2RuO_4$,\cite{li_NATPHYS_2025} similar studies of the $5d^4$ compounds could be a promising tool for breaking the $J=0$ state and inducing onset of local moments by transiently lowering the effective spin-orbit coupling below the critical value.

\begin{acknowledgments}
\end{acknowledgments}

\end{document}